\documentclass[aps,reprint,floatfix,superscriptaddress]{revtex4-1}

\usepackage[usenames,dvipsnames]{xcolor}

\definecolor{icolor1}{HTML}{0000FF}
\definecolor{icolor2}{HTML}{FFA400}
\definecolor{icolor3}{HTML}{008B00}
\definecolor{icolor4}{HTML}{FF34A7}
\definecolor{icolor5}{HTML}{632280}
\definecolor{icolor6}{HTML}{5E3100}

\newcommand{\SiV}{\textrm{SiV}}

\newcommand{\gev}{\textrm{GeV}}
\newcommand{\GeV}{\textrm{GeV}}

\usepackage[disable]{todonotes} 

\usepackage{graphicx}
\usepackage[separate-uncertainty=true, multi-part-units=single]{siunitx}
\usepackage{braket}
\newcommand*\autoref[1]{Figure \ref{#1}}

\begin{document}

\title{Optical and microwave control of germanium-vacancy center spins in diamond}

\author{Petr~Siyushev}
\email{petr.siyushev@uni-ulm.de}
\author{$\!^{, \dagger}$ Mathias~H.~Metsch}
\thanks{These two authors contributed equally}
\author{Aroosa~Ijaz}
\author{Jan~M.~Binder}
\affiliation{Institute for Quantum Optics, Ulm University, D-89081 Germany}
\author{Mihir~K.~Bhaskar}
\affiliation{Department of Physics,	Harvard University,	17 Oxford Street, Cambridge, MA 02138, USA}
\author{Denis~D.~Sukachev}
\affiliation{Department of Physics,	Harvard University,	17 Oxford Street, Cambridge, MA 02138, USA}
\affiliation{P. N. Lebedev Physical Institute of the RAS, Moscow 119991, Russia}
\author{Alp~Sipahigil}
\author{Ruffin~E.~Evans}
\author{Christian~T.~Nguyen}
\author{Mikhail~D.~Lukin}
\affiliation{Department of Physics,	Harvard University,	17 Oxford Street, Cambridge, MA 02138, USA}
\author{Philip~R.~Hemmer}
\affiliation{Electrical \& Computer Engineering Department, Texas A\&M University, College Station, TX 77843, USA}
\author{Yuri~N.~Palyanov}
\affiliation{Sobolev Institute of Geology and Mineralogy, Siberian Branch of Russian Academy of Sciences, Koptyug ave., 3, 630090, Novosibirsk, Russia}
\affiliation{Novosibirsk State University, 630090, Novosibirsk, Russia}
\author{Igor~N.~Kupriyanov}
\affiliation{Sobolev Institute of Geology and Mineralogy, Siberian Branch of Russian Academy of Sciences, Koptyug ave., 3, 630090, Novosibirsk, Russia}
\affiliation{Novosibirsk State University, 630090, Novosibirsk, Russia}
\author{Yuri~M.~Borzdov}
\affiliation{Sobolev Institute of Geology and Mineralogy, Siberian Branch of Russian Academy of Sciences, Koptyug ave., 3, 630090, Novosibirsk, Russia}
\affiliation{Novosibirsk State University, 630090, Novosibirsk, Russia}
\author{Lachlan~J.~Rogers}
\email{lachlan.j.rogers@quantum.diamonds}
\affiliation{Institute for Quantum Optics, Ulm University, D-89081 Germany}
\author{Fedor~Jelezko}
\affiliation{Institute for Quantum Optics, Ulm University, D-89081 Germany}
\affiliation{Center for Integrated Quantum Science and Technology (IQ$^\text{{st}}$), Ulm University, D-89081 Germany}

\begin{abstract}
A solid-state system combining a stable spin degree of freedom with an efficient optical interface is highly desirable as an element for integrated quantum optical and quantum information systems.
We demonstrate a bright color center in diamond with excellent optical properties and controllable electronic spin states. 
Specifically, we carry out detailed optical spectroscopy of a Germanium Vacancy ($\GeV$) color center demonstrating optical spectral stability.
Using an external magnetic field to lift the electronic spin degeneracy, we explore the spin degree of freedom as a controllable qubit.
Spin polarization is achieved using optical pumping, and a spin relaxation time in excess of \SI{20}{\micro\second} is demonstrated.
Optically detected magnetic resonance (ODMR) is observed in the presence of a resonant microwave field.
ODMR is used as a probe to measure the Autler-Townes effect in a microwave-optical double resonance experiment.
Superposition spin states were prepared using coherent population trapping, and a pure dephasing time of about \SI{19}{\nano\second} was observed.
Prospects for realizing coherent quantum registers based on optically controlled $\GeV$ centers are discussed. 
\end{abstract}


\maketitle


Over the last few decades significant effort has been directed towards the exploration of solid-state atom-like systems such as quantum dots or color centers in diamond owing to their potential application in quantum information processing \cite{
Gruber_S1997, 
santori2002indistinguishable, 
Lee_NN2013, 
muller2014optical}.  
The nitrogen vacancy (NV) center in diamond has become prominent due to its optical spin initialization and readout \cite{Jelezko_JPCM2004}, and the ease of spin control by microwave fields \cite{Gruber_S1997}.
However the small Debye-Waller factor of this defect~\cite{Jelezko_PSS2006} and its spectral instability~\cite{faraon2012coupling} hinder the realization of an efficient quantum-optical interface~\cite{nemoto2014network}, motivating an ongoing search for new candidates.
Here we investigate the recently discovered germanium vacancy ($\GeV$) center in diamond~\cite{iwasaki2015germanium-vacancy, palyanov2015germanium, ekimov2016germaniumvacancy}, demonstrating its outstanding spectral properties devoid of measurable spectral diffusion.
We show spin-$\frac{1}{2}$ Zeeman splitting which confirms this is the negative charge state of this defect.
We use two-photon resonance to optically prepare coherent dark spin superposition states, and show microwave spin manipulation via optically-detected magnetic resonance (ODMR).
The spin coherence time is found to be $T^\star_2=\SI{19(1)}{\nano\second}$, which is concluded to be limited by phonon-mediated orbital relaxation as in the closely-related silicon-vacancy ($\SiV$) center~\cite{rogers2014all-optical, jahnke2015electron-phonon}.
Optical and microwave control of $\GeV$ spin, combined with the possibility of $\GeV$ centers in nanophotonic devices~\cite{bhaskar2016quantum}, make it a promising platfrom for quantum optics and quantum information science applications.

\begin{figure*}
	\includegraphics[width=\textwidth]{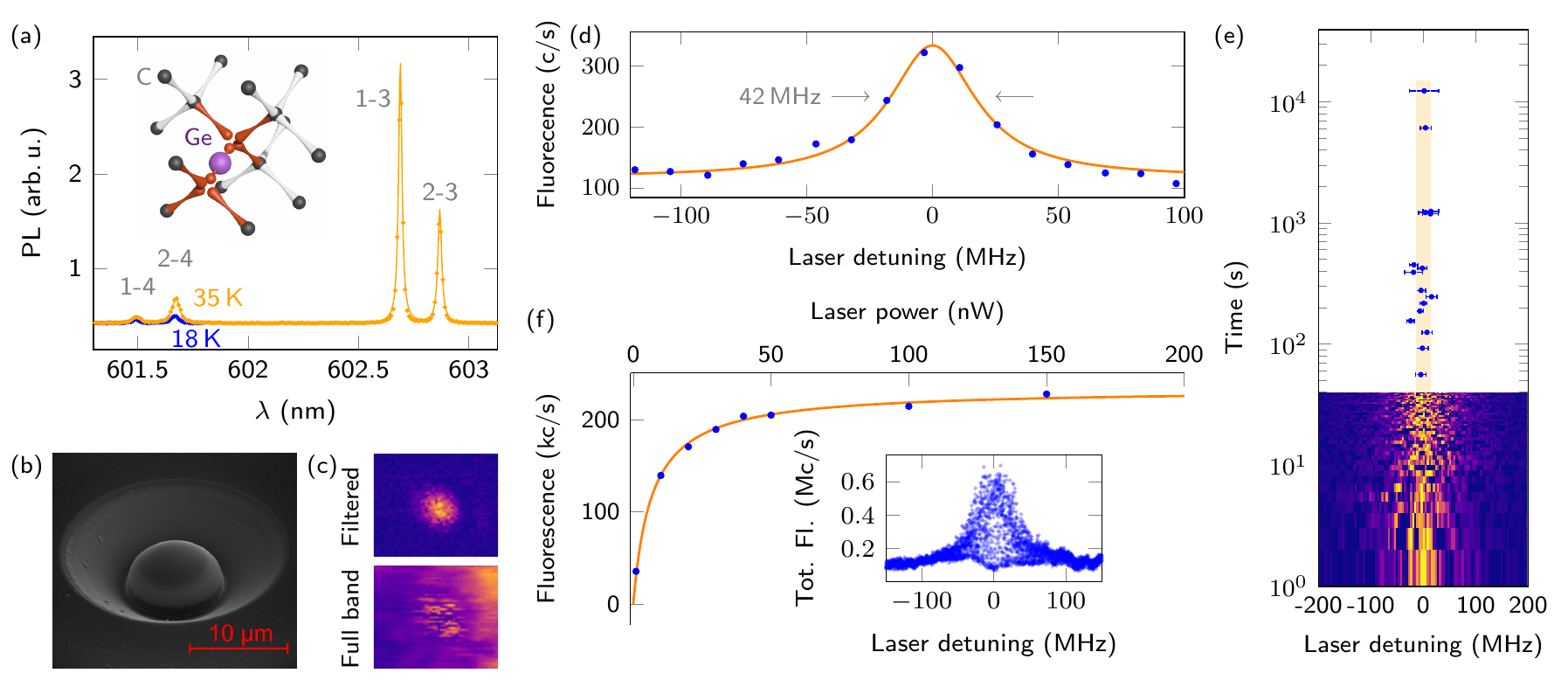}
	\caption{
		Stable narrow optical transitions.
		(a) PL spectrum of GeV at 18\,K and 35\,K, demonstrating the four-line ZPL structure with thermalization of the high-energy transitions. 
		(Inset) Molecular structure of the GeV center in diamond showing a Ge atom taking the place of two adjacent carbon atoms.
		(b) SEM image of a solid immersion lens fabricated on the sample surface.
		(c) Fluorescence image of a single $\GeV$ center located off-center under the SIL, recorded with a long-pass filter (sideband only) and without any filter on the detector.
		(d) In the low-intensity limit, linewidths as narrow as $\SI{42}{\mega\hertz}$ were recorded.
		(e) The $\GeV$ transitions were stable in frequency for measurements of several hours.
		(f) Saturation curve for a single GeV center recorded under resonant excitation (only photons from phonon side band are recorded). 
		The inset shows a single-sweep PLE spectrum recorded without long pass filter (entire emission band is recorded), yielding up to 0.6\,Mc/s at saturation intensity.
	}
	\label{fig:optical-props}
\end{figure*}

The $\GeV$ center can be produced in diamond during crystal growth and by ion implantation, and it fluoresces strongly with a zero-phonon line at 602\,nm accompanied by a weak phonon sideband (PSB) containing about 40\% of the fluorescence~\cite{iwasaki2015germanium-vacancy, palyanov2015germanium}.
Isotopic shifts of the fluorescence spectrum established that the defect contains a single germanium atom~\cite{ekimov2016germaniumvacancy, palyanov2016high-pressure}, and {\it ab initio} calculations suggest that it is formed by the Ge atom taking the place of two carbon atoms and relaxing its position to split the adjacent vacancy~\cite{Goss_PRB2005, iwasaki2015germanium-vacancy}.
The resulting geometry is aligned along a $\langle$111$\rangle$ axis and has inversion-symmetric D$_{3\mathrm{d}}$ symmetry as illustrated in \autoref{fig:optical-props}(a).
This structure is identical to the $\SiV$ center in diamond~\cite{hepp2014electronic, rogers2014electronic} and leads to $^{2}\mathrm{E}_\mathrm{g}$ and $^{2}\mathrm{E}_\mathrm{u}$ ground and excited states (respectively).
These have 2-fold degeneracy in both spin and orbit, which is partially lifted by spin-orbit interaction to produce levels labelled 1, 2 (ground) and 3, 4 (excited) in order of increasing energy.
Transitions between these lead to a four-line fine structure of the zero-phonon-line (ZPL) as shown in \autoref{fig:optical-props}(a)~\cite{palyanov2015germanium}.
At low temperatures the large spin-orbit splitting of the excited state results in thermal depopulation of level 4, apparent in the weaker 1-4 and 2-4 peaks relative to peaks 1-3 and 2-3 in the photoluminescence spectrum.

The experiments were performed on two identical samples polished in the \{100\} and \{111\} planes.
These diamonds were grown by high pressure high temperature synthesis in a Mg-Ge-C system, and the $\GeV$ centers were incorporated during this process~\cite{palyanov2015germanium, palyanov2016high-pressure}.
Solid immersion lenses were fabricated on the diamonds by focused ion beam milling (\autoref{fig:optical-props}(b)) to increase fluorescence collection efficiency~\cite{Hadden_APL2010,marseglia2011nanofabricated, castelletto2011diamond-based}.
Individual $\gev$ defects were excited resonantly by a tunable dye laser and addressed in a confocal microscope at cryogenic temperature ($T \approx \SI{2.2}{\kelvin}$) as shown in \autoref{fig:optical-props}(c).
An optical long pass filter with cut-on edge at \SI{610}{\nano\metre} was placed in the detection channel to reject scattered laser light, meaning that only photons associated with the phonon sideband were detected.
Measurements were controlled and coordinated using the Qudi software suite~\cite{binder2016qudi}.
%

%
A resonant laser was scanned across the transition between the lower branches of ground and excited state (transition $1\textnormal{-}3$) and a linewidth of \SI{42}{\mega\hertz} was measured as shown in \autoref{fig:optical-props}(d).
This is less than double the \SI{26}{\mega\hertz} transform-limit imposed by the excited state lifetime of \SI{6}{\nano\second}~\cite{bhaskar2016quantum}, indicating the stability of the transition over the measurement duration of \SI{40}{\second}.
Subsequent measurements at intervals over 4 hours demonstrated the line to be stable at longer time scales (\autoref{fig:optical-props}(e)).

The linewidth and stability were probed at lower laser excitation intensities to avoid power broadening.
\autoref{fig:optical-props}(f) presents a saturation measurement on the optical transition, yielding more than \SI{200}{\kilo c \per \second} detected in the sideband.
The off-center position of this $\GeV$ center in the SIL meant that laser scatter was collected only weakly, allowing a signal-to-noise ratio of 5:1 even after removing the filter from the detection path (\autoref{fig:optical-props}(c)).
Photoluminescence excitation (PLE) spectra are typically measured on a carefully filtered sideband to eliminate the dominant laser scatter.
The high contrast observed here without any filter is evidence of a strong optical transition.
Collecting the whole fluorescence band in this manner yielded up to \SI{0.6}{\mega c \per \second} as shown inset to \autoref{fig:optical-props}(f).
These data exhibit a blinking phenomenon that was observed to intensify for higher excitation intensities.
%
%
Blinking has also been observed for $\SiV$ centers in certain diamond samples \cite{neu2012photophysics, jantzen2016nanodiamonds} but not others \cite{rogers2014electronic}.
It is therefore anticipated that blinking is not intrinsic to the $\GeV$ center itself and can be controlled using superior sample preparation techniques.

\begin{figure}
	\includegraphics[width=\columnwidth]{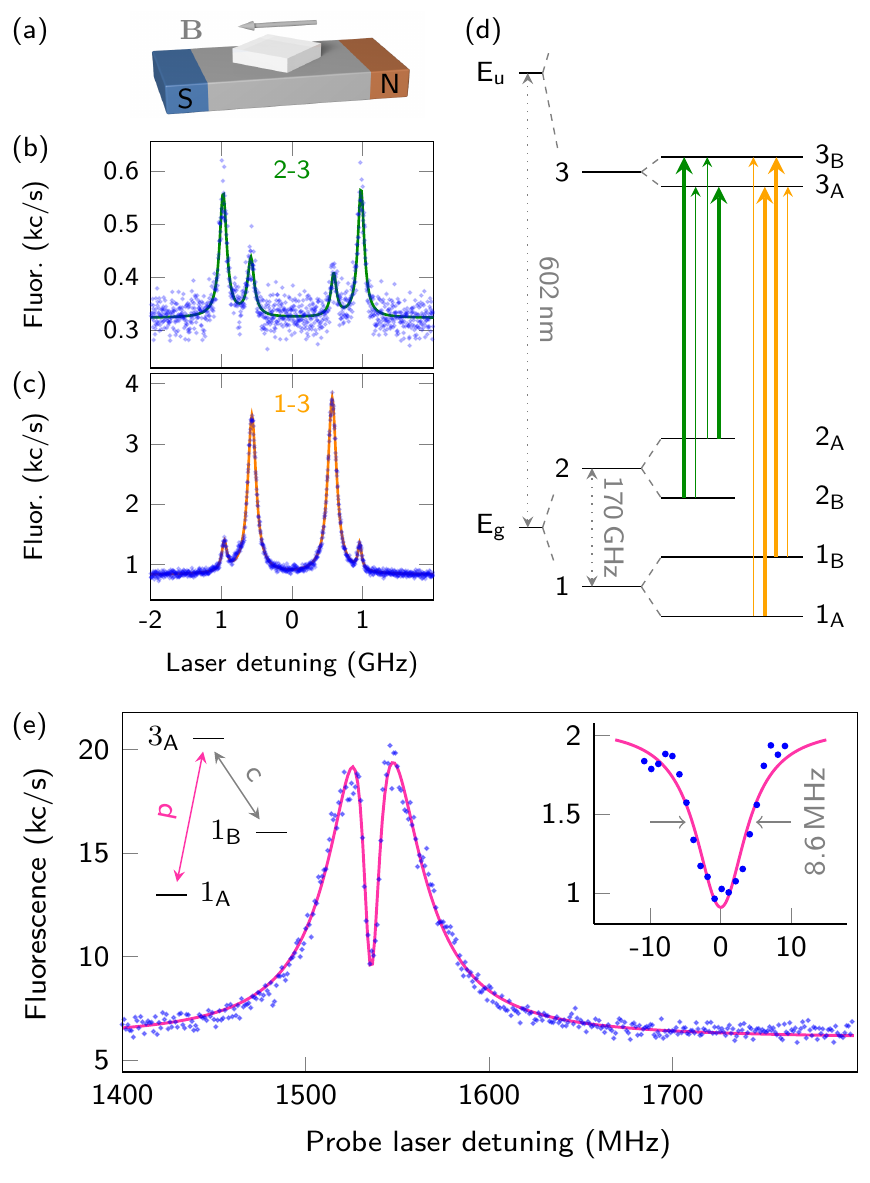}
	\caption{
		Optical lambda-schemes.
		(a) The diamond sample was mounted over neodymium magnets, giving a field in the plane of the \{100\} surface.
		(b) PLE spectrum shows transition 2-3 split into four peaks, with the outer two being strongest.
		(c) PLE spectrum shows transition 1-3 split into four peaks, with the inner two being strongest.
		(d) The magnetic field lifts the spin degeneracy and produces doublets from each of the ground and excited state branches (A and B subscripts describe two spin states).
		(e) CPT was performed on the $1_\textnormal{A}$-$3_\textnormal{A}$-$1_\textnormal{B}$ $\Lambda$-scheme, demonstrating coherent optical spin manipulation.
		Inset depicts the narrowest dip, which was found to be \SI{8.6(5)}{\mega\hertz} wide corresponding to a coherence lifetime of \SI{19(1)}{\nano\second}.
	}
	\label{fig:zeeman-spectra}
\end{figure}

The \{100\} sample was mounted over neodymium magnets producing a field of about 0.3\,T aligned roughly in the plane of the diamond surface as illustrated in \autoref{fig:zeeman-spectra}(a).
\autoref{fig:zeeman-spectra}(b, c) show the Zeeman-split PLE spectra for transitions 1-3 and 2-3 of an individual $\GeV$ center aligned almost perpendicular to the field.
For this center the ground state splitting between levels 1 and 2 was found to be about 170\,GHz, indicating the presence of transverse strain inducing an additional splitting of about 20\,GHz~\cite{bhaskar2016quantum}.
The measurement temperature of 2.2\,K was cold enough to reduce the thermal population in level 2 compared to level 1, and so the absolute intensity is less for transition 2-3 than for 1-3.
Transitions to state 4 are of less interest because of rapid thermalization into state 3~\cite{jahnke2015electron-phonon}, and were not measured here in PLE.
The four-fold Zeeman splitting patterns suggest electronic spin-$\frac{1}{2}$ splitting of each of the states, as illustrated in \autoref{fig:zeeman-spectra}(d). 
This is consistent with the 602\,nm band arising from the negative charge state of GeV.

Since optical transitions are spin-conserving, the stronger (weaker) peaks in \autoref{fig:zeeman-spectra}(b-c) correspond to transitions between states with more (less) similar spin projection.
For convenience we describe the two kinds of transitions as ``spin-conserving'' and ``spin-flipping'' respectively. 
The applied magnetic field component transverse to the $\langle111\rangle$ symmetry  axis leads to different spin quantization axes in the ground and excited states~\cite{muller2014optical}, resulting in the spin-flipping transitions being visible in the PLE spectra.
Interestingly, the conserving and flipping transitions have inverted order for transition $2\textnormal{-}3$ compared to $1\textnormal{-}3$.
This is in contrast to the observations for the $\SiV$ center~\cite{rogers2014all-optical} and warrants future investigation.
%
%

\begin{figure}
	\includegraphics[width=\columnwidth]{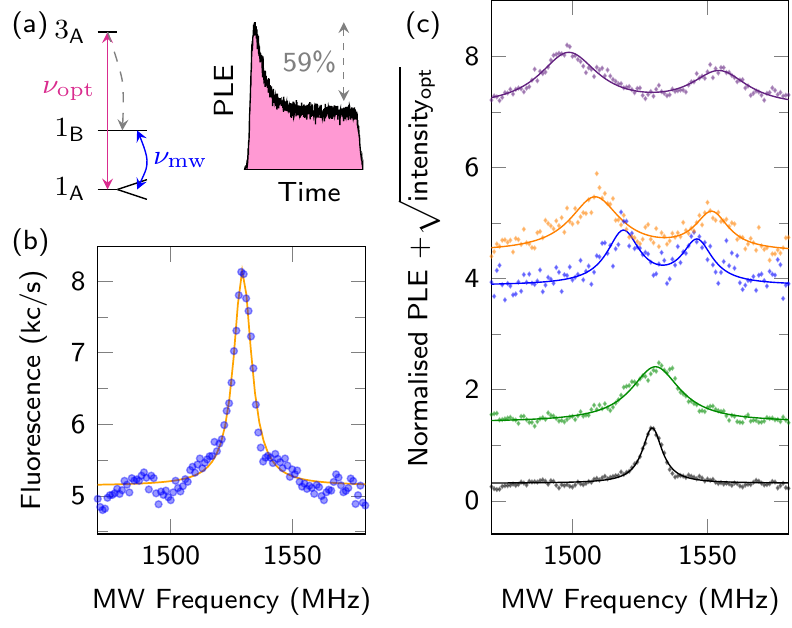}
	\caption{
		Microwave manipulation of $\GeV$ spin.
		(a) Excitation of the spin-conserving transition $1_\textnormal{A} \textnormal{-}3_\textnormal{A}$ polarizes the spin due to optical pumping into $1_\textnormal{B}$.
		The time-resolved fluorescence for an excitation pulse shows spin polarization of 59\% (limited by laser intensity).
		(b) Microwaves resonant to $1_\textnormal{A} \textnormal{-} 1_\textnormal{B}$ return population to $1_\textnormal{A}$ and restore the higher fluorescence, enabling optically detected magnetic resonance (ODMR).
		The width of the ODMR peak is \SI{9}{\mega\hertz} corresponding to a dephasing time of about $\SI{19}{\nano\second}$.
		The ODMR contrast of about 35\% is limited by the 59\% contrast possible from spin polarization.
		(c) Autler-Townes effect is observed upon laser power increase.
		The splitting of the ODMR signal scales linearly with the square root of laser intensity.
	}
	\label{fig:odmr}
\end{figure}

The presence of all four lines in the Zeeman-split spectrum of transition 1-3 indicates that lambda-schemes are accessible optically.
These provide an opportunity to investigate coherence properties through optically-prepared superpositions of the ground states \cite{fleischhauer2005electromagnetically}.
We make use of coherent population trapping (CPT), in which a dark superposition state is produced when driving optical fields are resonant to both transitions of the $\Lambda$-scheme.
The state is dark due to quantum interference, resulting in  a dip in the excitation spectrum with a width limited by the lifetime of the superposition.
Both excitation frequencies for CPT were generated from a single laser using a high-bandwidth electro-optic amplitude modulator with the carrier on transition $1_\textnormal{B} \textnormal{-} 3_\textnormal{A}$ as pump, and a sideband tuned across transition $1_\textnormal{A} \textnormal{-} 3_\textnormal{A}$ as probe.
The CPT linewidths in \autoref{fig:odmr}(b) are therefore insensitive to laser frequency noise, and the narrowest dip-width observed was \SI{8.6(5)}{\mega\hertz} corresponding to a coherence time of \SI{19(1)}{\nano\second}.

The optical $\Lambda$-schemes identified in the $\GeV$ electronic structure also provide a mechanism for polarizing the spin.
Resonantly exciting the spin-conserving transition $1_\textnormal{A} \textnormal{-} 3_\textnormal{A}$ led to optical pumping into the spin-B levels of the ground state as illustrated in \autoref{fig:odmr}(a).
Time-resolved fluorescence measurements indicate an optical pumping contrast of 59\%, which is the spin polarization (normalised population difference $(P_\textnormal{B} - P_\textnormal{A})/P_\textnormal{total}$) \cite{kirschner1979spin}.
This spin polarization is a function of the optical pumping rate compared to the spin relaxation rate, and can be increased with higher laser powers.
Since it is necessary to spectrally resolve the Zeeman-split transitions, excitation intensity (and hence spin polarization) is limited by the need to avoid power broadening beyond the Zeeman splitting.

A \SI{25}{\micro\metre} wire placed across the sample surface was used to apply microwaves at a frequency of \SI{1530}{\mega\hertz} corresponding to the $1_\textnormal{A} \textnormal{-} 1_\textnormal{B}$ transition, and a sharp increase in the steady-state fluorescence was observed (\autoref{fig:odmr}(b)).
This optically-detected magnetic resonance (ODMR) is a result of the microwaves returning population to the $1_\textnormal{A}$ level, and provides a simple direct way to manipulate the spin state.
Avoiding power broadening on the microwave and optical transitions produced an ODMR linewidth of \SI{9.0(5)}{\mega\hertz} (\autoref{fig:odmr}(b)), corresponding to a dephasing time of \SI{19(1)}{\nano\second}.
The ODMR contrast of about 35\% is limited by the spin polarization.
ODMR occurred at a microwave frequency exactly matching the two-photon detuning for laser fields producing CPT (\autoref{fig:zeeman-spectra}(e)), proving that the ODMR response is due to microwave manipulation of the ground state levels $1_\textnormal{A}$ and $1_\textnormal{B}$.
The ODMR linewidth and CPT dip width were also identical within the accuracy of the measurement, presenting a consistent picture regarding the coherence time of the Zeeman split levels.

Increasing the resonant excitation intensity on the optical transition $1_\textnormal{A} \textnormal{-} 3_\textnormal{A}$ led to a broadening and then splitting of the ODMR peak as shown in \autoref{fig:odmr}(c).
The splitting increased proportionally to the square root of laser intensity, confirming that it arises due to the optical Rabi frequency.
For optical Rabi oscillations faster than the excited state lifetime dressed states are resolvable, and the weak microwave field acts as a probe to measure this splitting.
This observation of the Autler-Townes effect indicates that coherent Rabi oscillations are achieved on the optical transition for excitation intensities above saturation.
It provides further evidence of the stable narrow optical transition for $\GeV$.

\begin{figure}
	\includegraphics[width=\columnwidth]{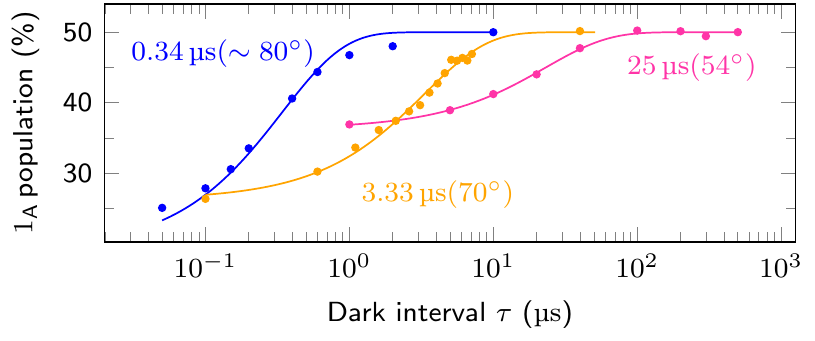}	
	\caption{
		Spin relaxation.
		Pulsed excitation was used to polarize the spin into $1_\textnormal{B}$ via optical pumping as in \autoref{fig:odmr}(a).
		The dark interval $\tau$ between laser pulses was varied to obtain the spin relaxation time (lines are exponential fits).
		For the $\sim80^\circ$ field misalignment in the ODMR and CPT measurements,  $T_1=\SI{0.34}{\micro\second}$ was measured.
		Improving the field alignment produced longer spin relaxation times ($\SI{3.33}{\micro\second}$ at $70^\circ$ and $\SI{25}{\micro\second}$ at $54^\circ$) but lower spin polarization.
		}
	\label{fig:spin-t1}
\end{figure}

Spin relaxation was measured by varying the dark intervals between the $1_\textnormal{A} \textnormal{-} 3_\textnormal{A}$ excitation pulses of \autoref{fig:odmr}(a), yielding $T_1=\SI{0.3}{\micro\second}$ as shown in \autoref{fig:spin-t1}(a).
Other field alignments were achieved by placing the sample flat on a pole of a disk magnet, producing a field normal to the sample surface.
For the \{111\} sample a number of $\GeV$ centers were measured to have $T_1=\SI{3.3(3)}{\micro\second}$.
The \{100\} sample gives a $\GeV$ misalignment of only $54^\circ$, and the spin relaxtion time was measured to be $T_1=\SI{25(5)}{\micro\second}$.
These results are shown in \autoref{fig:spin-t1}.
For $\SiV$ centres it was observed that spin $T_1$ extends for better aligned fields~\cite{rogers2014all-optical}, and the same phenomenon is exhibited here.
It is concluded that the $\GeV$ centers measured in the \{111\} sample had a misalignment of $70^\circ$,  and that the arbitrary field used for the ODMR and CPT measurements was misaligned by more than $80^\circ$.
Improved field alignment increases the spin relaxation time, but it also means the optical transition selection rules become more exclusive to the spin-conserving transitions.
It was therefore easier to achieve spin polarization for a misaligned field as shown by the equilibrium $1_\textnormal{A}$ population ($\tau\rightarrow0$) in \autoref{fig:spin-t1}, and this effect prevented a measurement of $T_1$ for the aligned $\GeV$ centres in the \{111\} sample.

The CPT and ODMR measurements indicate that levels $1_\textnormal{A}$ and $1_\textnormal{B}$ have a coherence time of about \SI{19}{\nano\second}, which is considerably shorter than the spin $T_1$.
The situation is most readily interpreted by analogy with the closely-related negatively-charged $\SiV$ center.
For $\SiV$ it was found that resonant phonons mediated orbital relaxation on a fast timescale of about \SI{40}{\nano\second} at 5\,K, and this limited the spin coherence time~\cite{rogers2014all-optical, jahnke2015electron-phonon, pingault2014all-optical}.
%
%
%
%
%
It seems that the same process occurs in the $\GeV$ center, however for similar phonon coupling parameters the orbital relaxation rate will be even faster as a result of the increased spin-orbit splitting in the ground state.
These energies are within the Debye approximation regime for diamond, and so the phonon density of states increases with energy.
This is offset by the more extreme thermal reduction in the transition rate out of level 1, but the measurements presented here suggest that the orbital $T_1$ lifetime is only about \SI{20}{\nano\second} at 2\,K even for level 1.
Although the spin $T_1$ increases with reduced transverse field, these phonon-mediated orbital transitions similarly limit the spin coherence time independent of the field alignment.
This picture is consistent with the temperature dependence of the transition linewidth and optical Rabi oscillation decay rate~\cite{bhaskar2016quantum}.

We have established that the $\GeV$ centre in diamond combines stable and bright optical transitions with an electronic spin-$\frac{1}{2}$ degree of freedom.
%
%
Even for a single centre this spin is accessible optically and by microwave fields, placing the $\GeV$ centre in a small class of color centres capable of ODMR.
ODMR is a powerful technique for access and manipulating solid-state spins \cite{Gruber_S1997}.
Broad interest in the NV center is based in part on its ODMR capabilities, and only a few other single color centres exhibit ODMR~\cite{Gruber_S1997,Koehl_N2011,Lee_NN2013,Siyushev_NC2014}.

The similarity of $\SiV$ and $\GeV$ physics indicates there is an entire family of quantum emitters in diamond with favorable optical properties.
We have shown that, like in the case of the $\SiV$, $\GeV$ spin coherence is limited by phonon dynamics that depend on temperature. 
Ongoing efforts to break this barrier for $\SiV$ by microstructures, sub-500\,mK temperature, and strain tuning, should also be beneficial for the $\GeV$ electron spin.
Combined with efficient optical access to the $\GeV$ electron spin reported here, such developments could enable large-scale quantum networks using $\GeV$ centers as quantum memory nodes. 
The demonstrated optical properties of the $\GeV$ center make this system worth persuing as a spin-photon interface.

%
%
%


The authors acknowledge funding from 
ERC, 
EU projects (SIQS, DIADEMS, EQUAM), 
DFG (FOR 1482, FOR 1493 and SFBTR 21), 
BMBF, 
USARL/ORISE, 
DARPA,
CUA,
ARL,
AFOSR,
NSF, 
the Volkswagen foundation
and Russian Science Foundation (grant no. 14-27-00054). 
%


PS and MM contributed equally to this work.
%
%
%

\listoftodos

\bibliography{paper_gev_spin}

\end{document}